\documentclass[english,hidelinks,preprint,aps,showpacs,superscriptaddress,nofootinbib]{revtex4-1}
\usepackage{bm}
\usepackage{amsmath}
\usepackage{amsthm}
\usepackage{amssymb}
\usepackage{stmaryrd}
\usepackage{mathbbol}   
\usepackage{newtxtext}
\usepackage{mathtools}
\usepackage{paralist}
\usepackage{color}

\usepackage{xcolor}
\definecolor{darkred}{rgb}{0.55, 0.0, 0.0}
\usepackage[unicode=true, 
 bookmarks=false,
 breaklinks=false,hidelinks=true,backref=false,colorlinks=true,linkcolor=darkred,citecolor=darkred]
 {hyperref}

\usepackage{tikz}
\usetikzlibrary{arrows.meta, positioning, calc}
\usetikzlibrary{decorations.pathreplacing,calligraphy}

\newtheorem{theorem}{Theorem}
\usepackage{siunitx}
\usetikzlibrary{angles,quotes}
\usetikzlibrary{patterns,snakes}
\tikzset{>=latex} 

\colorlet{xcol}{blue!85!black}
\colorlet{vcol}{green!60!black}
\colorlet{myred}{red!65!black}
\tikzstyle{vvec}=[->,vcol,thick,line cap=round]
\tikzstyle{ground}=[preaction={fill,top color=black!10,bottom color=black!5,shading angle=20},
                    fill,pattern=north east lines,draw=none,minimum width=0.3,minimum height=0.6]
\tikzstyle{metal}=[fill,top color=black!40,bottom color=black!20,shading angle=10]
\tikzstyle{mass}=[line width=0.6,black,fill=red!40!black!10,rounded corners=1,
                  top color=black!20,bottom color=black!10,shading angle=20]
\tikzstyle{pulcol}=[draw=blue!30!black,%fill=blue!40!black!10
                    top color=blue!40!black!20,bottom color=blue!40!black!10,shading angle=20]
\tikzstyle{rope}=[brown!70!black,very thick,line cap=round]
\def\rope#1{ \draw[black,line width=1.5] #1; \draw[rope] #1; }
\tikzstyle{mount}=[blue!20!black,fill,top color=blue!20!black!70,bottom color=blue!20!black!40,shading angle=10] %,line width=1.8,line cap=round
\tikzstyle{spring}=[line width=0.8,black!80,snake=coil,segment amplitude=5,segment length=5,line cap=round]
\pgfdeclarelayer{back} % to draw on background
\pgfsetlayers{back,main} % set order

\def\r{0.05} % pulley small radius
\tikzset{
  pics/pulley/.style={
    code={
      \draw[pulcol,line width=0.6] (0,0) circle (#1);
      \draw[pulcol,thick] (0,0) circle (\r);
  }},
  pics/mount/.style args={#1:#2}{ % angle, length
    code={
      \draw[mount] (0,0)++(#1-90:0.9*\r) arc (#1-90:#1-270:0.9*\r) --++ (#1:#2) --++ (#1-90:1.8*\r) -- cycle;
  }},
  pics/weight/.style args={#1,#2,#3}{ % bottom width, top width, height
    code={
      \draw[mass] (0,0) -- (#2/2,0) -- (#1/2,-0.7*#3)
        |- (-#1/2,-#3) -- (-#1/2,-0.7*#3) -- (-#2/2,0) -- cycle;
      \path[mass] (0,0) -- (0,-#3) node[pos=0.52] {$m$};
  }},
  pics/pulley/.default=0.3,
}

\usepackage[capitalize]{cleveref}
\usepackage{setspace}

\DeclareMathOperator*{\argmax}{arg\,max}

\def\ddd{\mathrm{D}}  % directional derivative

\def\tr{\mathrm{tr}\,}
\newcommand{\FE}[3]{\mathcal{F}_{#2,#3}(#1)} % nonequilibrium free energy
\newcommand{\FEeq}[2]{{F}^\mathrm{eq}_{#1,#2}} % equilibrium free energy
\def\pS{\rho}    % actual initial state
\def\qS{\sigma}  % optimal initial state
\def\ppeq{\pi} % generic Gibbs state

\def\pUnprep{\omega^{\circ}}
\def\pPost{\omega^\bullet}
\def\piS{\pi_0}
\def\piF{\pi_1}

\def\tempS{T_0}
\def\tempF{T_1}

\def\HUnprep{H^{\circ}}
\def\HPost{H^\bullet}
\def\HS{H_0}
\def\HF{H_1}

\def\stageI{A}
\def\stageII{B}
\def\stageIII{C}
\def\stageIV{D}

\def\map{\Phi}              % quantum channel
\def\Wmax{\mathcal{G}}      % free energy gain
\def\Wbase{G_\mathrm{base}}

\newcommand{\egap}{\epsilon}
\newcommand\Pjoint{P_\mathrm{tot}}

\allowdisplaybreaks
\begin{document}

\title{Maximizing free energy gain}

\author{Artemy Kolchinsky}
\affiliation{Universitat Pompeu Fabra}

\author{Iman Marvian}
\affiliation{Duke University Physics and Electrical Engineering}

\author{Can Gokler}
\affiliation{Harvard Engineering and Applied Sciences}

\author{Zi-Wen Liu}
\affiliation{Yau Mathematical Sciences Center, Tsinghua University}

\author{Peter Shor}
\affiliation{MIT Mathematics}

\author{Oles Shtanko}
\affiliation{IBM Quantum Almaden}

\author{Kevin Thompson}
\affiliation{Sandia National Laboratory}

\author{David Wolpert}
\affiliation{Santa Fe Institute}
\affiliation{Arizona State University}

\author{Seth Lloyd}
\thanks{To whom correspondence should be addressed: \url{slloyd@mit.edu}}
\affiliation{MIT Mechanical Engineering}

\begin{abstract}
Maximizing the amount of work harvested from an environment 
is important for a wide variety of biological and technological processes,
from energy-harvesting processes such as photosynthesis
to energy storage systems such as fuels and batteries.  
Here we consider the maximization of free energy --- and by extension, the maximum extractable work --- that can be gained by a classical or quantum system that undergoes driving by its environment. 
We consider how the free energy gain depends on the initial state of the system, while also accounting for the cost of preparing the system. 
We provide simple necessary and sufficient conditions for increasing the gain of free energy by varying the initial state. We also derive simple formulae that relate the free energy gained using the optimal initial state rather than another suboptimal initial state.  Finally, we demonstrate that the problem of finding the optimal initial state may have two distinct regimes, one easy and one difficult, depending on the temperatures used for preparation and work extraction. 
We illustrate our results
on a simple model of an information engine.
\end{abstract}
\maketitle

\section{Introduction} 

The last few decades have seen a revolution in non-equilibrium
statistical mechanics~\cite{jarzynski1997nonequilibrium,crooks1999entropy,crooks1998nonequilibrium,touchette2004information,seifert2012stochastic,parrondo2015thermodynamics}, 
with the realization that many
thermodynamic processes are governed by exact relations
such as the Jarzynski equality~\cite{jarzynski1997nonequilibrium} and the Crooks fluctuation theorem~\cite{crooks1999entropy}. 
An example of such a result
can be found in Ref.~\cite{kolchinsky2016dependence}, which derives expressions governing
the work dissipated by a system undergoing some driven process. Specifically, a simple formula is derived that relates the minimum amount of dissipated work, versus the actual amount dissipated, as a function of the initial statistical state of the process.

In this paper,  
we consider a physical system
that undergoes an interaction with its environment, 
as described by a driven classical or quantum-mechanical 
process.  
By extending the result mentioned above~\cite{kolchinsky2016dependence}, we calculate how much nonequilibrium free energy the system gains during this interaction. We also consider how the gain of free energy can be optimized by a judicious choice of the initial state. Optimizing the gain of free energy is physically meaningful, because the free energy gain sets a bound on the amount of work that the system can extract and store by interacting with a thermal environment: the maximum amount of work that can be extracted is equal to the nonequilibrium free energy minus the free energy at thermal equilibrium.

As a motivating example, consider a photosynthetic organism: before the sun rises in the morning, organism must invest resources in order to prepare its photosynthetic machinery for harvesting free energy from the sun. When the
sun sets in the evening, it stops photosynthesizing and uses
the harvested free energy to survive the night, reproduce, etc. 
All else being equal, the organism should prepare its photosynthetic machinery in the state that maximizes the gain of free energy, since this will typically translate into higher fitness.

In the next section, we formulate our general setup and use it to calculate the free energy gain as a function of the initial state. Importantly, our calculation takes into account the preparation of the initial state, as well as the extraction of free energy into a work reservoir. State preparation and work extraction may utilize external heat baths, possibly at two different temperatures. 

In our first result, we derive simple necessary and sufficient conditions to guarantee that, for a given interaction with the environment, free energy gain can be optimized by varying the initial state. 
We then derive a simple information-theoretic formula that describes the dependence of the free energy gain on the initial state. Using this formula, we relate the free energy gained when the process begins in an optimal initial state, versus that gained when the process begins in a suboptimal initial state. 
Finally, we show that the problem of identifying the optimal initial state exhibits two distinct regimes, depending on the bath temperatures involved in state preparation versus work extraction. When work extraction happens at a lower temperature  than preparation, the problem involves the maximization of a concave function, and it can be easily solved by gradient ascent. In this regime, 
a biological species in which each successive generation harvests more free energy 
is headed for the global maximum. On the other hand, when work extraction happens at a higher temperature than preparation, the objective may become nonconcave, and gradient ascent may get stuck in a suboptimal solution. At the end of this paper, we illustrate our results on an information engine.

Our results complement existing research on work extraction and free energy harvesting in classical and quantum thermodynamics~\cite{procaccia_potential_1976,esposito_second_2011,takara_generalization_2010,parrondo2015thermodynamics,shiraishi_quantum_2021,horodecki2013fundamental,brandao2013resource,sparaciari2017resource,allahverdyan2004maximal,skrzypczyk_work_2014}.  Such research typically considers how extractable work depends on properties of the physical process --- such as its speed of evolution~\cite{schmiedl2007optimal,nakazato2021geometrical}, constrained control~\cite{kolchinsky2021work,pinero2024optimization}, or stochastic fluctuations~\cite{solon2018phase,richens2016work} --- given some fixed initial state. Here we consider the complementary question of how extractable work depends on the initial state, given a fixed physical process. 
See also  Refs.~\cite{riechers2021initial,kolchinsky2021state,riechers2024thermodynamically,manzano2024thermodynamics,wolpert2023stochastic} for  related results concerning the dependence of entropy production on the initial state.

%``availability'', which refers to the maximum work that can be extracted from a fixed initial state given an optimized thermodynamic process~\cite{procaccia_potential_1976,esposito_second_2011,takara_generalization_2010,parrondo2015thermodynamics,shiraishi_quantum_2021,horodecki2013fundamental,brandao2013resource,sparaciari2017resource}.  Here we consider the complementary problem, concerning the amount of work that can be extracted from a fixed thermodynamic process, assuming one can optimize the initial state.  Our complementary perspective is related to research on ``thermodynamic capacity'' in quantum thermodynamics~\cite{faist_thermodynamic_2019}.

\section{Preliminaries}
\label{sec:prelim}

We consider a physical system that harvests free energy from its environment and extracts it as work.  The system may be classical or quantum, although for maximum generality, we usually employ quantum mechanical notation. For simplicity, we assume that the system is finite-dimensional, although most results can be extended to the infinite-dimensional case~\cite{kolchinsky2021state}. We will use the notation  $S(\rho)=-\tr\{\rho \ln \rho\}$ for the von Neumann entropy and $S(\rho \Vert \sigma)=\tr\{\rho (\ln \rho - \ln \sigma)\}$ for the quantum relative entropy.

Our analysis will use the relationship between work and free energy for isothermal processes. Consider a process that transforms some initial state and Hamiltonian $\rho,H$ to final state and Hamiltonian $\rho^\prime,H^\prime$, while connected to a heat bath at temperature $T$. According to the Second Law of Thermodynamics, the work that can be extracted during this transformation is bounded by the drop of \emph{nonequilibrium free energy}~\cite{procaccia_potential_1976,esposito_second_2011,takara_generalization_2010,parrondo2015thermodynamics,shiraishi_quantum_2021}:
\begin{align}
    W\le \FE{\rho}{H}{T} - \FE{\rho^\prime}{H^\prime}{T}\,,
    \label{eq:ff}
\end{align}
where nonequilibrium free energy is defined as
\begin{align}
    \FE{\rho}{H}{T} =\tr\{ \rho H \} - T \, S(\rho)\,.
    \label{eq:fedef0}
\end{align}
Throughout this paper, we choose energy units so that Boltzmann's constant is $k_B=1$. We also use the convention that $W>0$ indicates work extraction while $W<0$ indicates work investment. 

The bound~\eqref{eq:ff} can be achieved in a classical system using a slow (quasistatic) driving protocol that remains close to equilibrium throughout, and thus achieves thermodynamic reversibility~\cite{procaccia_potential_1976,esposito_second_2011,takara_generalization_2010}.  For quantum systems, this bound is achievable by a quasistatic protocol when the two states $\rho$ and $\rho^\prime$ are diagonal in their respective energy bases (as defined by $H$ and $H^\prime$ respectively)~\cite{muller_correlating_2018}.  
The bound is also achievable if the quasistatic protocol operates on a large number of identical copies of the quantum system, in which case \cref{eq:ff} refers to the work per copy. In the most general case, where $\rho$ and/or $\rho^\prime$ are non-diagonal and the protocol operates on a single copy of the system, the achievability of the bound remains an open question in quantum thermodynamics, possibly depending on available catalytic resources~\cite{lipka2024catalysis}.

With some rearrangement, the nonequilibrium free energy can also be expressed as 
\begin{align}
    \FE{\rho}{H}{T} =T \, S(\rho\Vert \ppeq) + \FEeq{H}{T}\,,
    \label{eq:fedef1}
\end{align}
where $\FEeq{H}{T}=\FE{\ppeq}{H}{T}$ is the equilibrium free energy, defined using the Gibbs state $\ppeq=e^{-H/T}/\tr\{e^{-H/T}\}$.   The first term, 
\begin{align}
T \,S(\rho\Vert \ppeq) =\FE{\rho}{H}{T}-\FEeq{H}{T} \,,\label{eq:availabilitydef}
\end{align}
is called the \emph{availability}. It quantifies the maximum work that can be extracted from the state $\pS$ by bringing it to equilibrium $\ppeq$.

\section{Free energy harvesting}

Suppose that the system has access to an internal work reservoir (e.g., a battery), which is used for state preparation and work extraction. Suppose also that the system also has access to two heat baths, at temperatures $\tempS$ and $\tempF$. The system undergoes the following four-stage procedure, also illustrated in Figure~\ref{fig:stages}:  
\smallskip

\begin{figure}
\begin{tikzpicture}[
    block/.style={circle, draw, inner sep=0pt, fill=red!70, minimum size=1.2cm, text=white},
    blueblock/.style={circle, draw, inner sep=0pt,fill=blue!70, minimum size=1.2cm, font=\large, text=white},
    arrow/.style={-Latex, thick},
    textarrow/.style={pos=0.5, sloped, above, font=\small, rotate=20},
    label/.style={font=\small, rotate=90}
]
\tikzstyle{bath}=[draw=blue!40!black,top color=blue!10,
                                  bottom color=blue!20,shading angle=30,thick,rounded corners=1]
\tikzstyle{source}=[draw=red!50!black,top color=red!20,
                                   bottom color=red!30,shading angle=30,thick,rounded corners=1]

% Nodes

\def\bottomA{-2}
\def\topA{4}

\node (rho0) at (-0.5,3) {$\pS,\HS$};
\node (phi_rho0) at (5,\topA) {$\map(\pS),\HF$};
\node (pi0) at (-1.5,\bottomA+1) {$\pUnprep,\HUnprep$};
\node (pi1) at (6.5,\bottomA) {$\pPost,\HPost$};

% Circles for T0 and T1
%\node[block] (T0) at (0,1.5) {$\tempS$};
%\node[blueblock] (T1) at (5,1.5) {$\tempF$};

%\draw[red!25,line width=10pt]  (-1.5,1.5) --  (0,1.5) ;

% Arrows
\draw[arrow] (rho0) -- (phi_rho0) node[midway, above, font=\small,sloped] {\stageII. Interaction};
\draw[arrow] (pi1) -- (pi0) node[midway, below, font=\small,sloped] {\stageIV. Reset};
\draw[arrow] (pi0) -- (rho0) node[above,midway, font=\small,sloped] {\stageI. Preparation};
\draw[arrow] (phi_rho0) -- (pi1) node[above,midway,font=\small,sloped] {\stageIII. Work Extraction};

\draw[arrow]  (-5,\bottomA) -- (-5,\topA) node[midway, above, sloped] {Free energy};

\draw[source]
(0.5,0.5) rectangle ++(-1.2,1) node[midway,align=center] {$\tempS$};
\draw[bath]
(5.45,0.5) rectangle ++(-1.2,1) node[midway,align=center] {$\tempF$};

\def\h{0.6}  % mass height
\def\w{0.8}  % mass width
\def\R{0.3}  % pulley radius

\def\W{1.8}  % ground width
\def\H{2.3}  % ground height
\def\L{0.2}  % rope length
\def\R{0.30} % pulley radius
\def\px{-1.5*\W} % pulley x position
\def\py{0.70*\H} % pulley y position
\def\my{0.40*\H} % mass y position
\def\rx{\px-1.0*\R} % rope-peg x position

\rope{ %draw[rope,line cap=round]
(\rx,0.3*\H) -- (\px-\R,\py) arc(180:0:\R) --++ (0,\my-\py)}
\pic at (\px,\py) {pulley={\R}};
\draw[mass] (\px+\R,\my)++(-\w/2,0) rectangle++ (\w,-\h) node[midway] {};

\def\px{6.8} % pulley x position
\rope{ %draw[rope,line cap=round]
(\rx,0.3*\H) -- (\px-\R,\py) arc(180:0:\R) --++ (0,\my-\py)}
\pic at (\px,\py) {pulley={\R}};
\draw[mass] (\px+\R,\my)++(-\w/2,0) rectangle++ (\w,-\h) node[midway] {};
    \end{tikzpicture}

\caption{\label{fig:stages}
Four-stage protocol used to harvest free energy from the environment. During the Preparation stage, the system is coupled to the internal work reservoir and a heat bath at temperature $\tempS$. During Interaction, the system harvests free energy from the external environment. During Work Extraction, the system is coupled to the internal work reservoir and a heat bath at temperature $\tempF$. During Reset, the system is again coupled to the external environment.
}
\end{figure}

\noindent {(\stageI)} \emph{Preparation}: the system begins in an unprepared state $\pUnprep$ and Hamiltonian $\HUnprep$. It is then 
driven to the prepared state $\pS$ and Hamiltonian $\HS$, while coupled to the internal work reservoir and heat bath at temperature $\tempS$. 
Given \cref{eq:ff}, the work extracted during this transformation is bounded by 
\begin{align}
W_\text{\stageI} \le \FE{\pUnprep}{\HUnprep}{\tempS}-\FE{\pS}{\HS}{\tempS} \,.
\label{eq:a0}
\end{align}

\smallskip
\noindent {(\stageII)} \emph{Interaction/Free energy harvesting}:  
the system is disconnected from the work reservoir. It then undergoes a fixed interaction with the environment, which may contain any number of thermodynamic reservoirs, free energy sources, and external work reservoirs (e.g., the sun).   
At the end of this stage, the system has Hamiltonian $\HF$ and state $\map(\pS)$. Here $\map$ is the quantum channel (completely positive and trace-preserving map) that describes the system's evolution due to the interaction with the environment.  

\smallskip
\noindent {(\stageIII)} \emph{Work Extraction}: the system is coupled to the work reservoir and the heat bath at temperature $\tempF$. It is then driven from state $\map(\pS)$ and Hamiltonian $\HF$ to final state $\pPost$ and Hamiltonian $\HPost$. 
According to \cref{eq:ff}, the maximum work that can be extracted during this transformation is  
\begin{align}
W_\text{\stageIII} \le  \FE{\map(\pS)}{\HF}{\tempF}-\FE{\pPost}{\HPost}{\tempF} \,.
\label{eq:b0}
\end{align}

\smallskip
\noindent {(\stageIV)} \emph{Reset}: the system is disconnected from the internal work reservoir and then undergoes another interaction with the environment. As a result of this interaction --- which, in some cases, may be a simple relaxation --- the system ends in state $\pUnprep$ and Hamiltonian $\HUnprep$. 
This completes the cycle, thereby preparing the system for Stage {\stageI}. 
In the special case where $\pPost=\pUnprep$ and $\HPost=\HUnprep$, the Reset stage is not necessary. %and the equilibrium free energies are equal ($\FSeq=\FFeq$).

As a concrete --- though still highly idealized --- example of our setup, one might imagine a simple photosynthetic system, such as Bacteriorhodopsin in archaea~\cite{lanyi2004bacteriorhodopsin,pinero2024optimization}. During Preparation, the organism spends free energy (by hydrolyzing ATP) to synthesize the Bacteriorhodopsin protein from free-floating amino acids. During Interaction, the protein uses solar energy to pump protons across the cellular membrane, thereby increasing the membrane potential. During Work Extraction, the membrane potential is used by ATPase to synthesize ATP. Reset may occur by consumption of any additional ATP and degradation of the Bacteriorhodopsin protein back into amino acids.  We note that in this system, Preparation and Work Extraction steps may occur at different temperatures.

We now calculate a bound on the work that can be extracted using this four-stage process. Since the system only interacts with the work reservoir during Stages {\stageI} and {\stageIII}, the total amount of extracted work  is
\begin{align}
    W=W_\text{\stageIII}+W_\text{\stageI}.
\end{align}
\cref{eq:a0,eq:b0} then imply the upper bound $W  \le \Wmax(\pS)$, where we define 
\begin{align}
\Wmax(\pS)=[\FE{\map(\pS)}{\HF}{\tempF}-\FE{\pS}{\HS}{\tempS}]-[\FE{\pPost}{\HPost}{\tempF} - \FE{\pUnprep}{\HUnprep}{\tempS}] \,.
\label{eq:deltaF}
\end{align}
Observe that $\Wmax(\pS)$ is a function of the initial state and that it consists of two terms. The first term is the gain of nonequilibrium free energy during the Interaction with the environment (Stage \stageII{}). 
The second term is the loss of nonequilibrium free energy during the Reset (Stage \stageIV{}). 

We may also rewrite $\Wmax(\pS)$ in the following form:
\begin{align}
\Wmax(\pS) &=\tempF  \, S(\map(\pS) \Vert \piF) - \tempS \, S( \pS \Vert \piS) + \Wbase \,,
\label{eq:deltaA}
\end{align}
where $\piS=e^{-\HS/\tempS}/\mathrm{tr}\{e^{-\HS/\tempS}\}$ and $\piF=e^{-\HF/\tempF}/\mathrm{tr}\{e^{-\HF/\tempF}\}$ refer to Gibbs states corresponding to $(\HS,\tempS)$ and $(\HF,\tempF)$ respectively. 
This expression follows by combining \cref{eq:fedef1,eq:deltaF} and rearranging, while also defining the ``baseline'' term
\begin{align}
\Wbase &= \big[\FE{\pUnprep}{\HUnprep}{\tempS}-\FEeq{\HS}{\tempS}\big] -\big[\FE{\pPost}{\HPost}{\tempF} - \FEeq{\HF}{\tempF}\big]
\label{eq:wbase}
\end{align}
Observe that \cref{eq:deltaA} expresses $\Wmax(\pS)$ as the gain of availability, \cref{eq:availabilitydef}, during the transition from $\pS$ to $\map(\pS$), plus a constant offset ($\Wbase$).

In the following, we will generally be interested in the dependence of $\Wmax(\pS)$ on the initial state $\pS$. In \cref{eq:deltaF}, this dependence is captured by first term, the gain of nonequilibrium free energy, since the second term does not depend on $\pS$ (nor on the channel $\map$). In \cref{eq:deltaA}, this dependence is captured entirely by the gain of availability, since $\Wbase$ again does not depend on $\pS$ (nor on the channel $\map$). 

For convenience, we will often refer to $\Wmax(\pS)$ simply as the ``free energy gain''.

\section{Increasing free energy gain}
\label{sec:optgain}

We now consider the problem of maximizing the free energy gain $\Wmax(\pS)$ with respect to the initial state $\pS$. Before proceeding, we note that maximizing free energy gain is not always the same as maximizing extracted work $W$. The two optimization problems are equivalent when the Preparation and Work Extraction stages are thermodynamically reversible, so that the bounds~\eqref{eq:a0} and \eqref{eq:b0} are saturated.  %As discussed in Section~\ref{sec:prelim} above, typically this is achievable when Preparation and Work Extraction stages can be carried out by quasistatic driving protocols. 
The optimization of free energy gain $\Wmax(\pS)$ is relevant when Preparation and Work Extraction stages are thermodynamically optimal. Moreover, because $\Wmax(\pS)$ always sets an upper bound on extractable work, maximizing $\Wmax(\pS)$ is also relevant when the precise details of Preparation and Work Extraction are unknown, varying, or simply undefined.

To study the problem of optimizing  $\Wmax(\pS)$, we first consider a few special cases. First, suppose that $\map$ is a (generalized) Gibbs-preserving map that transforms the initial Gibbs state to the final Gibbs state $\piF$, $\map(\piS) = \piF$.   
This situation applies if the driving by the environment is sufficiently slow so that the system remains in equilibrium throughout (quasistatic driving), or if the system is allowed to equilibrate at the end of its interaction with the environment. Then, by monotonicity of relative entropy~\cite{muller-hermes_monotonicity_2017}, 
\begin{align}
S(\map (\pS) \Vert \piF)=S(\map (\pS) \Vert \map (\piS)) \le S(\pS \Vert \piS)\,.
\label{eq:klineq}
\end{align}
Combining with \cref{eq:deltaA} implies that $\Wmax(\pS) \le \Wbase$ for all $\pS$ whenever $\tempF \leq \tempS$.  
That is, if $\map$ maps $\piS$ to $\piF$ and the 
temperature of Work extraction is less than or equal to the temperature of Preparation, it is impossible to extract more work than $\Wbase$, regardless of the initial state. Moreover, since $\Wmax(\piS)=\Wbase$, one cannot do better than the naive strategy of setting the initial state to $\piS$, i.e., letting the system relax fully to equilibrium for Hamiltonian $\HS$ and temperature $\tempS$.

On the other hand, suppose that $\map$ is not Gibbs preserving, so $\map(\piS)\ne\piF$.  Suppose we still choose the initial state as $\piS$.  \cref{eq:deltaA} then gives
\begin{align}
    \Wmax(\piS) = \tempF  \, S(\map(\piS) \Vert \piF) + \Wbase > \Wbase \,,
    \label{eq:wbasebound}
\end{align}
where the last inequality follows from the positivity of the relative entropy between different states. 
Thus, if $\map$ is not Gibbs preserving, there is always at least one initial state for which $\Wmax$ is strictly greater than $\Wbase$. Moreover, generally $\Wmax$ can be increased even further by optimizing the choice of the initial state.

These conditions are formalized by the following statement.

\begin{theorem}\label{thm:1}~\\
(1a) If $\map(\piS)=\piF$, there exists some $\pS$ with $\Wmax(\pS)>\Wbase$ only if $\tempF > \tempS$.

\noindent (1b) If $\map(\piS)\ne\piF$, there always exists some $\pS$ with $\Wmax(\pS)>\Wbase$.
\end{theorem}

\smallskip

As a special case, consider the situation where the quantum channel is the identity ($\map(\rho)=\rho$ for all inputs) and the Hamiltonians $\HS,\HF$ are equal. In that case, $\map(\piS)=\piF$ only if $\tempS = \tempF$. Then, Theorem~\ref{thm:1} implies that free energy can be gained beyond $\Wbase$ if and only if $\tempS \ne \tempF$. In this special case, the interaction with the environment provides no free energy, so free energy can only be gained using a temperature difference between the baths,  as in a heat engine.

\section{Dependence on the initial state}
We now consider how the free energy gain depends on the choice of the initial state.  Before proceeding, we provide a useful expression
for the (one-sided) directional derivative of $\Wmax$. Recall that the directional
derivative  at state $\qS$ toward state $\pS$ is defined
as
\begin{equation}
\ddd_{\pS-\qS} \Wmax({\qS})=\lim_{\lambda\to0^{+}}\frac{
\Wmax[\qS+\lambda(\pS-\qS)]-\Wmax(\qS)
}{\lambda}
\label{eq:dd}
\end{equation}
In Appendix~\ref{app:ddd}, we show that  this directional derivative can be expressed as
\begin{align}
\ddd_{\pS-\qS} \Wmax({\qS})=\Wmax(\pS)-\Wmax(\qS)+\tempS\, S(\pS\Vert\qS)-\tempF\,S[\map(\pS)\Vert\map(\qS)] \,.
\label{eq:prop1}
\end{align}
This expression is particularly useful when considering $\qS$ for which the directional derivatives vanishes. This is shown in the following result, which is proved in Appendix~\ref{app:proofs}.
\begin{theorem}
\label{thm:dd}Let $\qS$ be a (local or global) minimum, maximum,
or saddle point of $\Wmax$. Then, for any $\pS$ with $S(\pS\Vert\qS)<\infty$,
\begin{align}
\Wmax(\qS)-\Wmax(\pS)=\tempS\,S(\pS\Vert\qS)-\tempF\,S[\map(\pS)\Vert\map(\qS)].
\label{eq:thmddres}
\end{align}
\end{theorem}

Theorem~\ref{thm:dd} means that the increase in free energy gain when 
using initial state $\qS$ versus $\pS$ has a universal information-theoretic
expression. %, given by contraction of the temperature-scaled relative entropies. 
% This expression depends only on two states ($\qS$ and
% $\pS$), the quantum channel ($\map$), and the two temperatures ($\tempS$
% and $\tempF$). Importantly, there is no dependence on the equilibrium distributions $\piS,\piF$, nor on the baseline contribution $\Wbase$. 
While the left-hand side of \cref{eq:thmddres} contains thermodynamic terms, the right-hand side of this equality consists purely of information-theoretic 
quantities, that is relative entropies, scaled by the temperatures.  
The change in the relative entropies 
can be understood as the loss of distinguishability between $\pS$ and $\qS$ during the 
process, and it does not explicitly depend on the energy functions. 
Indeed, \cref{thm:dd} provides a simple 
example of a relationship between information-theoretic and physical quantities. 
Such relationships have
been found to be very useful in the resource theory of  thermodynamics~\cite{horodecki2013fundamental,Brandao2015,guryanova2016thermodynamics,sparaciari2017resource}.

\section{Optimal initial state}

We now consider the problem of optimizing the initial state so as to maximize free energy gain.  Consider the initial state $\qS$ that is a local or global maximizer of $\Wmax$. Then, according to \cref{thm:dd}, for any other initial state $\pS$ with $S(\pS\Vert\qS)<\infty$,
\begin{align}
\label{eq:dd3}
\Wmax(\qS)-\Wmax(\pS)=\tempS\,S(\pS\Vert\qS)-\tempF\,S[\map(\pS)\Vert\map(\qS)]
\end{align}
 \cref{eq:dd3} provides a simple formula for computing 
the free energy gain that is lost when the system is prepared 
in the ``wrong'' initial state: it 
is given by comparing the
relative entropy between the ``wrong'' and ``right'' (i.e., optimal)
states at the beginning of the process with the
the same relative entropy at the end of the process, multiplied by the temperatures $\tempS$ and $\tempF$.

Next, we consider the difficulty of finding the optimal state $\qS$. As we now show, $\Wmax$ is a concave function of the initial state, if the temperature of Preparation is no cooler than the temperature of Work Extraction.  The proof is found in \cref{app:proofs}.

\begin{theorem}
\label{thm:concavity}
$\Wmax(\pS)$ is a concave function of $\pS$ if $\tempS\ge\tempF$.
\end{theorem}

Any local maximizer of a concave function must also be a global maximizer. Thus, \cref{thm:concavity} implies that as long as $\tempS\ge\tempF$, global optimization of free energy harvesting 
can be done by a simple procedure, e.g., gradient ascent in the space of density matrices~\cite{riechers2024thermodynamically}. 
Consider an adaptive system that undergoes the
same free-energy harvesting process many times.    Each time the system goes through the free energy harvesting cycle, 
it has the opportunity to vary its initial state $\pS$ to try
to increase the free energy gain.  The concavity of free energy maximization implies that if the adaptive process is able to alter the initial state to improve the amount of free energy harvested in each round, then the adaptive system is headed for the global optimum, and will not get stuck in a local optimum.

A population of photosynthetic bacteria, for example, exhibits genetic variation in the individuals' molecular mechanisms for performing the preparation, free energy harvesting, and work extraction stages, all of which impact the viability of the individual organisms and their ability to reproduce.   In general, we expect that the more efficient an individual bacterium is at harvesting free energy, the more viable it will be, resulting in its offspring forming a larger fraction of the population in subsequent generations.  Focusing only on the free energy harvesting stage, we see that genetic variations which increase the amount of free energy harvested -- e.g., a small change in the structure of a photo-harvesting chromophore which provides greater overlap with the absorptive spectrum of the chromophore and ambient light conditions -- will guide the population as a whole to adapt its composition to become more efficient at energy harvesting.

The concavity of free energy gain as a function of the initial state of the bacteria implies that if the free energy harvesting is suboptimal, then there is always a nearby initial state that improves the free energy gain.  The only way for the adaptive process to become stuck in a local optimum is if genetic variation in the population is unable to explore fully the space of initial probabilistic states.

If Work Extraction occurs at a warmer temperature than Preparation, however, in general there is no guarantee that $\Wmax$ is concave. In this case, finding the global optimum may be a much harder problem, and gradient ascent may get trapped in suboptimal local maxima.

Finally, we note that \cref{eq:dd3} only holds for those $\pS$ that obey $S(\pS\Vert\qS)<\infty$. This condition is equivalent to the requirement that the support of $\pS$ falls within the support of $\qS$. For this reason, the applicability of \cref{eq:dd3}  is most general in those cases where the optimizer $\qS$ has full support. In our final result, we provide some simple sufficient conditions
for $\qS$ to have full support. The proof is found in Appendix~\ref{app:proofs}. 
\begin{theorem}
Any (local or global) maximizer $\qS$ of $\Wmax$ has full support if $\tempS>0$ and $\map(\rho)$
has full support for all $\rho$, or if $\tempS>\tempF$. 
\label{thm:fsupport}
\end{theorem}

\section{Example}

\def\pScl{p_X}
\def\qScl{q_X}
\def\pEnv{p_Y}
\def\pXY{p_{XY}}
\def\pScl{p}
\def\qScl{q}
\def\pEnv{p_\mathrm{env}}
\def\pXY{p_\mathrm{tot}}
\def\HSys{H}
\def\HEnv{H_\mathrm{env}}
\def\HXY{H_\mathrm{tot}}

\def\HHH{\mathcal{H}}

We illustrate our results using a simple example of a two-level system. The system can be considered as a model of an ``information engine'' which gains free energy by interacting with a heat bath and a low-entropy environment~\citep{mandal2012work,barato2013autonomous}. 

For simplicity, we begin by focusing on the classical case, where all Hamiltonians, states, and channels are diagonal in the same basis. We use classical notation $\pScl$ (instead of $\pS$) to indicate the actual initial probability distribution of the system, $\qScl$ (instead of $\qS$) to indicate the optimal initial distribution, and $P$ (instead of $\map$) to indicate the classical transition matrix (conditional probability of outputs given inputs). We also use the notation $D(\pScl\Vert \qScl)=\sum_x \pScl(x) \ln [{\pScl(x)}/{\qScl(x)}]$ to indicate the classical relative entropy (also known as Kullback-Leibler divergence) and $\HHH(\pScl)=-\sum_x \pScl(x) \ln {\pScl(x)}$ for the Shannon entropy.

The engine is modeled as an overdamped two-level system $X\in\{0,1\}$ with energy gap $\egap\ge 0$, with energy function $\HSys(0)=0,\HSys(1) = \egap$. 
The engine is coupled to the \emph{environment}, another two-level  system $Y\in\{0,1\}$ with a uniform energy function, $\HEnv(0)=\HEnv(1) = 0$. 
The engine and environment are weakly-coupled, so their joint energy function can be decomposed as $\HXY({x,y}) \approx  \HSys(x)+\HEnv(y)$. 
Also, initially at time $t=0$,  the engine and the environment are statistically independent: $\pXY(x,y)=\pScl(x)\pEnv(y)$.  
Then, over the time interval $t\in[0,\tau]$, the two systems relax freely while coupled to a heat bath at temperature $T=1$. The environment and the engine have coupled transitions: the $0\rightarrow1$  transition in the engine
occurs only when the environment simultaneously undergoes a $0\rightarrow1$ transition, and vice versa for the $1\rightarrow 0$ transition. No transitions occur in/out of microstates where the engine and environment occupy different levels, $(x,y)\in \{(0,1),(1,0)\}$.

Suppose the engine is used to extract work using the 4 stage protocol shown in Figure~\ref{fig:stages}.  We assume that Stage \stageI{} (Preparation)  starts and ends with the same Hamiltonian ($\HUnprep=\HS=\HSys$), and that the unprepared state is the Gibbs state $\piS$ at some temperature $\tempS$ ($\pUnprep=\piS$).   During Stage \stageII{} (Interaction), the engine evolves according to a transition matrix $P$, defined below. 
Finally, we assume Stage \stageIII{} (Work Extraction) starts and ends with the same Hamiltonian ($\HF=\HPost=\HSys$) and that the final state is the Gibbs state $\piF$ at some temperature $\tempF$. 

The net amount of extracted work is bounded $W\le \Wmax(\pScl)$ by the gain of availability:
\begin{align}
    \Wmax(\pScl)= \tempF  \, D(P \pScl \Vert \piF) - \tempS \, D( \pScl \Vert \piS)
    \label{eq:ex0}
\end{align}
This result follows from \cref{eq:deltaA} and the fact that $\Wbase=0$ (given our assumptions). 
At the same time, our results imply that $\Wmax$ can be expressed as
\begin{align}
\Wmax(\pScl)=\Wmax(\qScl)-\big[\tempS\,D(\pScl \Vert \qScl)-\tempF\,D(P \pScl\Vert P \qScl)\big]\,,
    \label{eq:ex1}
\end{align}
where $\qScl\in\argmax_{\pScl} \Wmax(\pScl)$ is a maximizer of $\Wmax$, as given \cref{eq:dd3}, which is valid whenever $\qScl$ has full support. A simple sufficient condition for $\qScl$ to have full support is for environment distribution $\pEnv$ to have full support; this follows from  \cref{thm:fsupport} and because then $P \pScl$ has full support for all $\pScl$ (see \cref{eq:infoengineT} below). The function $\Wmax$ and the optimizer $\qScl$ will depend on the parameters of the problem: the energy gap $\egap$, the initial environment distribution $\pEnv$, and the temperature of Preparation ($\tempS$), Interaction ($T$), and Work Extraction ($\tempF$).

To construct $P$, we assume that the engine and environment undergo continuous-time Markovian dynamics, represented by the rate matrix  
\begin{align*}
R =\begin{bmatrix}-1 & 0 & 0 & e^{\egap}\\
0 & 0 & 0 & 0\\
0 & 0 & 0 & 0\\
1 & 0 & 0 & -e^{\egap}
\end{bmatrix}\,,
% \label{eq:R}
\end{align*}
where $R_{ij}$ is the transition rate from state $j$ to state $i$, with the four states referring to  $(x,y)=\{(0,0)$, $(0,1)$, $(1,0),(1,1)\}$.  $R$ obeys local detailed balance for the energy function $\HXY$ and interaction temperature $T=1$. 
We consider the limit of a long relaxation, corresponding to the following joint transition matrix: 
\begin{align}
\Pjoint = \lim_{\tau\rightarrow\infty}e^{\tau R} 
=\begin{bmatrix}\frac{1}{1+e^{-\egap}} & 0 & 0 & \frac{1}{1+e^{-\egap}}\\
0 & 1 & 0 & 0\\
0 & 0 & 1 & 0\\
\frac{e^{-\egap}}{1+e^{-\egap}} & 0 & 0 & \frac{e^{-\egap}}{1+e^{-\egap}}
\end{bmatrix}\,.\label{eq:T}
\end{align}
The transition matrix of the engine subsystem $X$ is computed by 
 marginalizing, 
\begin{align}
 P(x^\prime \vert x)  = \sum_{y,y^\prime} \Pjoint(x^\prime ,y^\prime \vert x,y) \pXY(y\vert x) = \sum_{y,y^\prime} \Pjoint(x^\prime ,y^\prime \vert x,y)\pEnv(y)\,,
\end{align}
and it can written explicitly  in matrix notation as
\begin{align}
P=\frac{1}{1+e^{-\egap}}\begin{bmatrix}
1+e^{-\egap}-\pEnv(0) e^{-\egap} & 1-\pEnv(0) \\
\pEnv(0) e^{-\egap} & \pEnv(0)+e^{-\egap}
\end{bmatrix}
\label{eq:infoengineT}
\end{align}

\begin{figure}
\includegraphics[width=\textwidth]{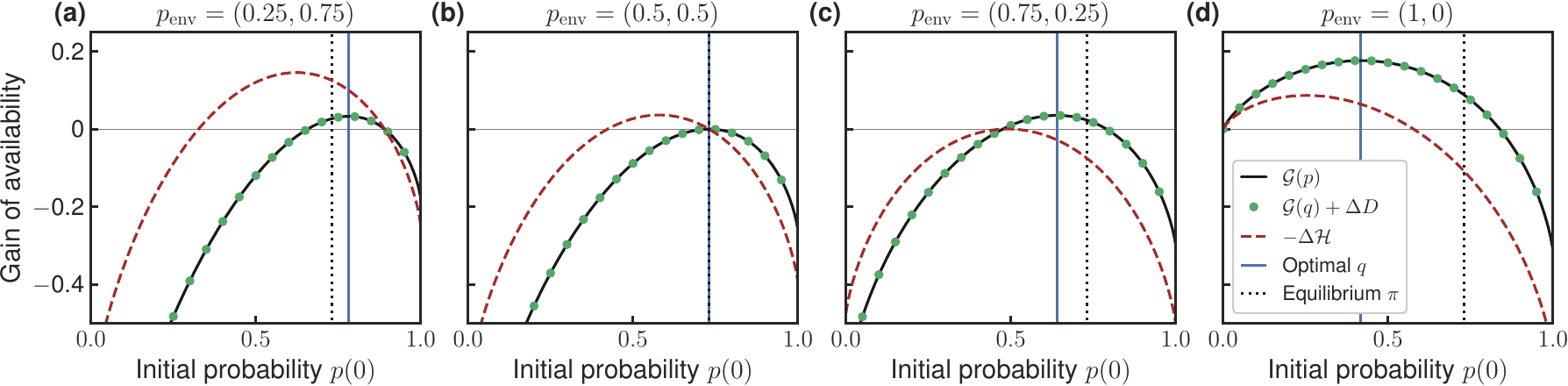}
\caption{\label{fig:newexample1}
Availability gain $\Wmax(\pScl)$ as a function of the engine initial distribution $(p(0),1-p(0))$. \textbf{(a)}-\textbf{(d)} correspond to four different environment initial distributions $\pEnv$. Black lines show $\Wmax(\pScl)$ computed using \cref{eq:ex0}; 
green dots indicate predictions made using our information-theoretic expression~\eqref{eq:ex1} (using shorthand $\Wmax(\qScl)+\Delta D$ in legend). 
Optimal initial distribution $\qScl$ and equilibrium initial distribution $\pi$ are indicated using vertical lines.  Dashed curve indicates reduction of the engine's Shannon entropy as a function of initial distribution, $-\Delta \HHH$ from \cref{eq:deltaHHH}. Vertical axes have the same scale. 
Other parameters: $\tempS=\tempF=T=1$, $\egap=1$.
}
\end{figure}

We now illustrate our results with some numerical experiments. 
In Figure~\ref{fig:newexample1}, we show the gain of availability $\Wmax(p)$ as a function of the engine's initial distribution $\pScl$, computed using \cref{eq:ex0}.   Since the engine only has two microstates, $\pScl$ is fixed by probability $\pScl(0)$ of microstate $X=0$, shown on the horizontal axis.  The different subplots correspond to different initial distribution of the environment $\pEnv$. The other parameters are set to 
$\egap=1$ and $\tempS=\tempF=T=1$.  
We show the location of the optimal initial distribution $\qScl$ found by numerical optimization, and the value of $\Wmax(\pScl)$ is computed using using our information-theoretic expression~\eqref{eq:ex0}.  We also show the location of the initial equilibrium distribution $\piS=\big(\frac{1}{1+e^{-\egap}},\frac{e^{-\egap}}{1+e^{-\egap}}\big)$.

We comment on some interesting aspects of the results in \cref{fig:newexample1}. 
First,  $\Wmax$ is a concave function of the initial distribution, in accordance with \cref{thm:concavity}. We also verify our main result, showing that the thermodynamic~\eqref{eq:ex0} and information-theoretic~\eqref{eq:ex1} expressions for $\Wmax$ are equivalent.
Note that maximum availability gain is non-monotonic in $\pEnv(0)$. In \cref{fig:newexample1}(b), the environment starts from the maximum entropy state $\pEnv=(0.5,0.5)$; in this case, the optimal initial distribution is the equilibrium one ($\qScl=\piS$), and it is not possible to harvest strictly positive availability ($\Wmax(\qScl)=0$). On the other hand, strictly positive availability can be harvested when the environment is biased to microstate  $Y=0$, see \cref{fig:newexample1}(a), or $Y=1$, see \cref{fig:newexample1}(c)-(d), but the optimal strategy differs in these two cases. When the environment is biased to $Y=1$ ($\pEnv(0)< 0.5$), the optimal initial distribution is biased to $X=0$  relative to equilibrium ($\qScl(0)> \pi(0)$). Conversely, when the environment is biased to $Y=0$ ($\pEnv(0)< 0.5$), the optimal initial distribution is biased toward $X=1$  relative to equilibrium ($\qScl(0)< \pi(0)$). This reflects the balance between two effects. On one hand, there is an advantage to biasing the engine's initial distribution toward $X=0$, because the transition $0\to 1$ harvests $\egap$ energy from the heat bath. On the other hand, availability can also be harvested by decreasing the Shannon entropy of the engine, i.e., by increasing
\begin{align}
    -\Delta \HHH := \tempS \HHH(p)-\tempF \HHH(Pp) \,.\label{eq:deltaHHH}
\end{align}
This quantity is shown as the dashed red line in \cref{fig:newexample1}. It can be seen that this second effect shift the optimal distribution toward $X=1$ relative to  equilibrium, and it becomes stronger when the environment is more concentrated on state $Y=1$.

\cref{fig:newexample1}(d)  shows that $\qScl$ has full support even though $\pEnv=(1,0)$ does not have full support (so $P\pScl$ does not have full support for all $\pScl$; see \cref{eq:infoengineT}). This demonstrates that \cref{thm:fsupport} is only a sufficient, but not necessary, condition for the optimizer $\qScl$ to have full support.

\begin{figure}
\includegraphics[width=\textwidth]{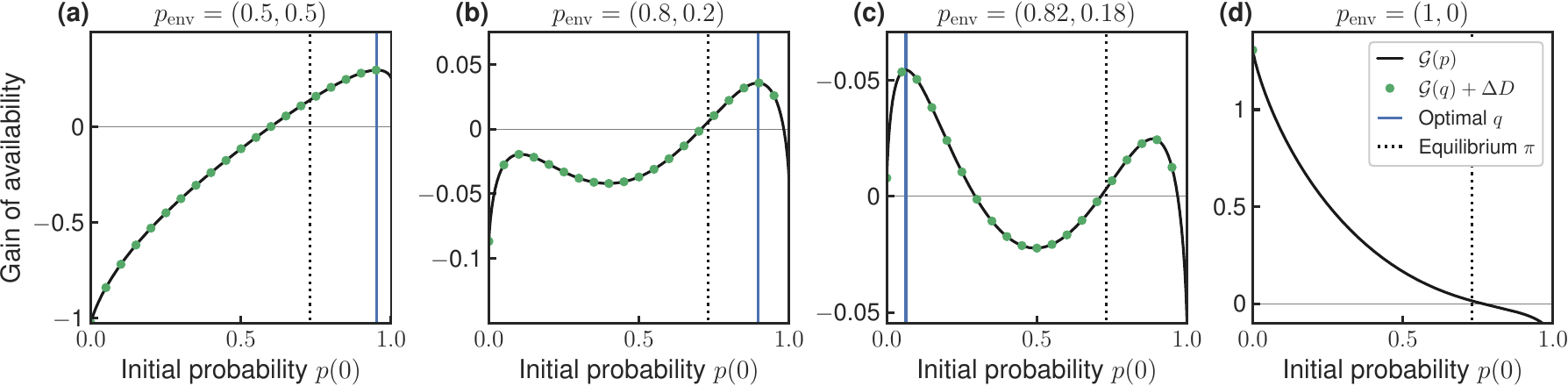}
\caption{\label{fig:newexample2}
Same as in  \cref{fig:newexample1}, but where temperature of Work Extraction is higher than of Preparation and Interaction, $\tempF=3 > \tempS = 1$. Black lines show $\Wmax(\pScl)$ computed using \cref{eq:ex0}; 
green dots indicate predictions made using information-theoretic expression~\eqref{eq:ex1}. 
\textbf{(a-d)} correspond to different initial states of the environment. Observe that in some cases, the function $\Wmax$ is non-concave and may have  multiple local maxima.  In \textbf{(d)}, the optimal distribution $q$ does not have full support, so the equivalence between Equations~\eqref{eq:ex0} and \eqref{eq:ex1} does not hold.}
\end{figure}

In  \cref{fig:newexample2}, we consider the same system but now setting  the temperature of Work Extraction  higher than that of Preparation,  $\tempS=T=1,\tempF=3$. As above, different subplots correspond to  different initial distributions of the environment. To emphasize interesting features, we make several changes with respect to
\cref{fig:newexample1}: we explore different initial environment distributions, the scale of the y-axes are different, and for simplicity we do not show  the change of Shannon entropy $-\Delta \HHH$.)

When $\tempF > \tempS$, \cref{thm:concavity} no longer applies and the function $\Wmax(\pScl)$ may become non-concave, as seen in \cref{fig:newexample2}(b)-(d). Moreover, \cref{fig:newexample2}(b)-(c) show that $\Wmax$ may even have multiple local maxima. Note that in these two plots, the shape of $\Wmax$ changes and the identity of the higher local maxima switches, even though $\pEnv$ undergoes a very small change.  In Figure \cref{fig:newexample2}a-c, we verify that the thermodynamic~\eqref{eq:ex0} and information-theoretic~\eqref{eq:ex1} expressions for $\Wmax$ are equivalent. 
For systems with several local maxima/minima, we verified that this equivalence holds regardless of which critical point is chosen as $\qScl$.

Note that $\tempF > \tempS$ is necessary but not sufficient for non-concavity, since $\Wmax$ remains concave in \cref{fig:newexample2}(a). Also, \cref{fig:newexample2}(d), $\pEnv$ does not have full support and \cref{thm:fsupport} no longer holds. In this case, the optimal initial distribution $\qScl=(0,1)$ does not have full support, so our information-theoretic expression no longer applies.

\begin{figure}
\includegraphics[width=\textwidth]{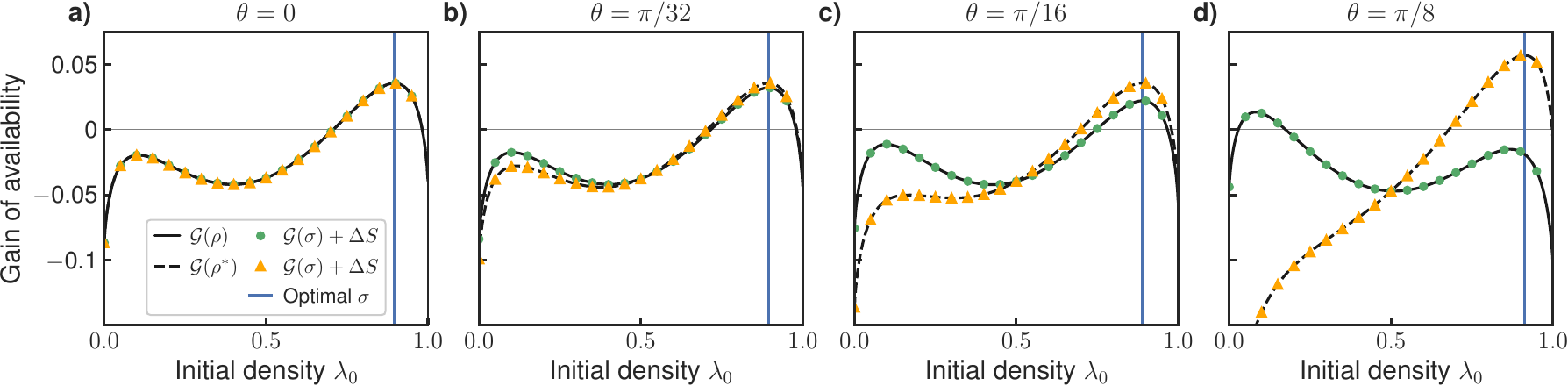}
\caption{\label{fig:newexampleQ}
Gain of availability $\Wmax(\rho)$ in a quantum system for different amounts of coherence (parameterized by $\theta$). Solid black line shows $\Wmax$ for states diagonal in the reference basis,  dashed black line shows $\Wmax$ for states diagonal in the basis of the optimizer $\sigma$, both calculated using \cref{eq:ex0q}. Markers indicate predicted values of $\Wmax$ from information-theoretic expression~\eqref{eq:ex1q}. 
\textbf{(a)} For $\theta=0$ (no coherence), we recover the classical result shown in \cref{fig:newexample2}b. 
\textbf{(b)-(d)} Advantage of selecting initial state in the optimal basis increases with increased coherence. Vertical axes have the same scale. See text for details.}
\end{figure}

In our last numerical experiment, we consider this system in the quantum regime.  We define a quantum channel $\map$ that dephases any input state $\rho$ in the reference basis $\{\vert 0\rangle,\vert 1\rangle\}$, then applies the transition matrix $P$ in this basis:
\begin{align}
\Phi(\rho) =\sum_{x\in\{0,1\},x^\prime\in\{0,1\}} P_{x^\prime x} \vert x^\prime \rangle \langle x^\prime \vert \langle x\vert \rho \vert x\rangle.
\end{align}
The Hamiltonians used for Preparation and Work Extraction are allowed to be coherent (non-diagonal) with respect to the reference basis. Specifically, we consider the same Hamiltonian with energy gap $\egap$, but rotated by angle $\theta$ with respect to the reference basis: 
\begin{align}
    \HUnprep=\HS=\HF=\HPost= \egap U_\theta \vert 1\rangle\langle 1\vert U^\dagger_\theta 
    \qquad \qquad 
    U_\theta = 
\begin{bmatrix}
\cos\theta & {-\sin}\theta \\
\sin\theta & \cos\theta
\end{bmatrix}
\end{align}
The net amount of extracted work is bounded by the gain of availability:
\begin{align}
    \Wmax(\rho)= \tempF  \, S(\Phi (\rho) \Vert \piF) - \tempS \, S( \pScl \Vert \piS)
    \label{eq:ex0q}
\end{align}
where $\piS$ and $\piF$ indicate Gibbs states for $(H_0,T_0)$ and $(H_1,T_1)$ respectively. As above, our results imply that $\Wmax$ can also be expressed as
\begin{align}
\Wmax(\rho)=\Wmax(\sigma)-\big[\tempS\,S(\rho \Vert \sigma)-\tempF\,S(\Phi(\rho) \Vert \Phi(\sigma))\big]\,,
    \label{eq:ex1q}
\end{align}
where $\sigma\in\argmax_{\rho} \Wmax(\rho)$ is a maximizer of $\Wmax$. \cref{eq:ex0q,eq:ex1q} are simply the quantum versions of \cref{eq:ex0,eq:ex1}.

\cref{fig:newexampleQ} shows the results for  
four values of $\theta$, which varies the amount of coherence. The other parameters are chosen as in \cref{fig:newexample2}(b) ($\tempS=T=1,\tempF=3,\egap=1,\pEnv=(0.8,0.2)$).    
For each $\theta$, we use \cref{eq:ex0q} to calculate the value of $\Wmax$ for two sets of states $\rho$, always indexed by the (lower energy) eigenvalue $\lambda_0$. First, solid lines indicate $\Wmax$ for states $\rho$ diagonal in the reference basis, $\rho = \lambda_0\vert 0\rangle \langle 0\vert+(1-\lambda_0)\vert 1\rangle \langle 1\vert$.  Second, dashed lines indicate $\Wmax$   for states $\rho^*$ diagonal in the same basis as the optimal initial state $\sigma$.  Vertical lines indicate the location of $\sigma$ in this optimal basis. We also plot the values of $\Wmax$ calculated using our information-theoretic expression \cref{eq:ex1q}, verifying that it matches those calculated using \cref{eq:ex0q} for both sets of states.

There is no coherence in \cref{fig:newexampleQ}(a), so we effectively recover the classical case shown in \cref{fig:newexample2}(b). For higher $\theta>0$, 
\cref{fig:newexampleQ}(b)-(d) demonstrate that availability gain can be increased by preparing the initial state in the correct basis.   Moreover, we verified that the optimal state $\sigma$ is not diagonal in either the reference basis nor the basis of the rotated Hamiltonian $\HS=\HF$. The optimal basis arises from the balance of two effects: the cost of preparation (which favors $\sigma$ being in the same basis as $H$) versus the  free energy dissipated when $\sigma$ is dephased by $\Phi$ (which favors $\sigma$ being in the reference basis). Under the optimal strategy, the engine gains availability not only by increasing energy or decreasing entropy, but also by harvesting coherence with respect to the basis of $\HS=\HF$.

\section{Discussion}

In this paper, we investigated how the gain of free energy depends on the initial state, while considering a broad class of classical and quantum process. We first 
showed that the initial state can be used to optimize the gain of free energy if and only if a process fails to map equilibrium distributions to equilibrium distributions. 
We also derived information-theoretic formulae
for the gain of free energy as a function of the initial state, and we then used these to quantify the difference in free energy gain
between the optimal initial state and some suboptimal initial state.
This difference was shown to be equal to the drop of the relative entropy between the initial and final states, scaled by temperatures.

For macroscopic systems, the deficit in free energy harvested by a suboptimal
initial state may itself be a macroscopic quantity.   
Moreover, for a living system that requires free energy to survive and 
reproduce, there is considerable evolutionary pressure
to increase the amount of free energy harvested. 
The difficulty of maximizing this objective depends on whether it is concave or not. We derived conditions for the free energy gain to be a concave function of the initial state. When these conditions hold, the objective can be maximized using a simple strategy like gradient ascent; for example, a species where each generation does slightly better
at harvesting free energy will eventually approach the global maximum. 
In cases where the conditions do not hold, the free energy gain may become nonconcave. In such cases, the maximization of the objective becomes qualitatively more difficult, and simple strategies like gradient ascent may become trapped in local optima.

As mentioned in Section~\ref{sec:optgain} above, the optimization of free energy gain is not necessarily the same as optimization of extracted work $W$, although the two problems become equivalent when Preparation and Work Extraction stages are thermodynamically reversible.  
When the Preparation and Work Extraction stages are not reversible, the state that maximizes extracted work ---  for example, as might be found varying some control parameters of Preparation stage (assuming all else held fixed) --- is not necessarily the same as the state that maximizes free energy gain.  An interesting direction for future research would consider optimization of extracted work, assuming some realistic constraints on Preparation and/or Work Extraction protocols, e.g., constraints on preparable initial states or finite-time constraints.

\vspace{10pt}
\vspace{10pt}

\begin{acknowledgements}This work was supported by NSF under an
INSPIRE program.  S.L. was supported by ARO and AFOSR.
AK was partly supported by the European Union's Horizon 2020 research and innovation programme under the Marie  Sk{\l}odowska-Curie Grant Agreement No. 101068029, and by Grant 62828 from the John Templeton Foundation. AK and DHW would like to thank the Santa Fe Institute for helping 
to support this research.  This paper was also made possible through 
the support of Grant No. TWCF0079/AB47 from the Templeton 
World Charity Foundation, Grant No. FQXi-RFP-1622 from the 
FQXi foundation, and Grant No. CHE-1648973 from the U.S. National 
Science Foundation. 
The opinions expressed in this paper 
are those of the authors and do not necessarily 
reflect the view of Templeton World Charity Foundation or the John Templeton Foundation. 
\end{acknowledgements}

\appendix

\section{Derivation of directional derivative, Eq.~\eqref{eq:prop1}}
\label{app:ddd}
Using \cref{eq:deltaA} and \cref{eq:dd}, we may write the directional derivative of $\Wmax$ as
\begin{align}
\ddd_{\pS-\qS}\Wmax({\qS})=\tempF \,\ddd_{\map(\pS)-\map(\qS)}S(\map(\qS)\Vert\piF) -\tempS\, \ddd_{\pS-\qS}S(\qS\Vert\piS)\,,\label{eq:dd00}
\end{align}
where the directional derivatives of the relative entropy terms are with respect to the first argument. 
The directional derivative of the
relative entropy $S(\qS\Vert\piS)$ is~\cite[Lemma~1]{audenaertContinuityBoundsQuantum2005} 
\begin{align*}
\ddd_{\pS-\qS} S(\qS\Vert\piS)  = \mathrm{tr}\{(\pS-\qS)(\ln\qS-\ln\piS)\} \,.
\end{align*}
Rearranging the right hand side gives
\begin{align*}
\ddd_{\pS-\qS}S(\qS\Vert\piS) & =S(\pS\Vert\piS)-S(\pS\Vert\qS)-S(\qS\Vert\piS)\,.
\end{align*}
Similarly, the directional derivative of the relative entropy $S(\cdot\Vert\piF)$ obeys
\begin{align*}
\ddd_{\pS-\qS}S(\map(\qS)\Vert\piF) & =S[\map(\pS)\Vert\piF]-S[\map(\pS)\Vert\map(\qS)]-S[\map(\qS)\Vert\piF]\,.
\end{align*}
Plugging these expressions into \cref{eq:dd00} gives  
\begin{align*}
\ddd_{\pS-\qS}\Wmax({\qS})&=\tempF\big[S[\map(\qS)\Vert\piF]-S[\map(\pS)\Vert\map(\qS)]-S[\map(\pS)\Vert\piF]\big]\\
&\qquad-\tempS[S(\pS\Vert\piS)-S(\pS\Vert\qS)-S(\qS\Vert\piS)].
\end{align*}
Finally, we plug in the definitions
of $\Wmax(\qS)$ and $\Wmax(\pS)$ from \cref{eq:deltaA} and rearrange to give \cref{eq:prop1}.

\section{Proofs}
\label{app:proofs}

\begin{proof}[Proof of \cref{thm:dd}]
If $S(\pS\Vert\qS)<\infty$, then the support of $\pS$ falls within the support of $\qS$, which in turn implies that $\qS-\alpha\pS$ is positive-definite
for some $\alpha>0$~\cite[p.~15]{watrous2020advanced}. Therefore, it is possible to move along the line $\qS + \lambda (\pS-\qS)$ both toward $\pS$ ($\lambda>0$) and away from $\pS$ ($\lambda < 0$). If $\qS$ is a minimum (or maximum), the directional derivative $\ddd_{\pS-\qS} \Wmax({\qS})$ must
vanish, since otherwise $\Wmax$ could be decreased (or increased) by a small
perturbation toward or away from $\pS$. On the other hand, if $\qS$ is a saddle point, the directional derivative
vanishes by definition. The result follows by plugging $\ddd_{\pS-\qS}\Wmax(\qS)=0$ into 
\cref{eq:prop1} and rearranging.
\end{proof}

\begin{proof}[Proof of \cref{thm:concavity}]
For any pair of states $\pS$ and $\qS$, let $\qS_{\lambda}:=(1-\lambda)\qS+\lambda\pS$
refer to the convex mixture with weight $\lambda\in[0,1]$. Next, we define the following quantity:
\begin{align*}
\chi & =(1-\lambda)\Wmax(\qS)+\lambda\Wmax(\pS)-\Wmax(\qS_{\lambda})\,,
\end{align*}
which measures the degree of convexity $\Wmax$ over the states $\{\qS_{\lambda}:\lambda\in[0,1]\}$. It is positive when $\Wmax$ is convex, negative when concave, and zero when linear.  Using \cref{eq:deltaA}, we may express $\chi$
as 
\begin{align}
\begin{aligned}
\chi & =\tempF[(1-\lambda)S(\map(\qS)\Vert\piF)+\lambda S(\map(\pS)\Vert\piF)-S(\map(\qS_{\lambda})\Vert\piF)]\\
 & \qquad-\tempS[(1-\lambda)S(\qS\Vert\piS)+\lambda S(\pS\Vert\piS)-S(\qS_{\lambda}\Vert\piS)]
\end{aligned}
\label{eq:Www}
\end{align}
The bracketed term in second line can be rearranged as
\begin{align*}
  (1-\lambda)S(\qS\Vert\pi_{0})+\lambda S(\pS\Vert\pi_{0})-S(\qS_{\lambda}\Vert\pi_{0}) =(1-\lambda)S(\qS\Vert\qS_{\lambda})+\lambda S(\pS\Vert\qS_{\lambda})
\end{align*}
In a similar way, the bracketed term in the first line can be rearranged as 
\begin{align*}
(1-\lambda)S(\map(\qS)\Vert\map(\qS_{\lambda}))+\lambda S(\map(\pS)\Vert\map(\qS_{\lambda}))
\end{align*}
Plugging back into \cref{eq:Www} gives 
\begin{align*}
\chi & =(1-\lambda)[\tempF\,S(\map(\qS)\Vert\map(\qS_{\lambda}))-\tempS S(\qS\Vert\qS_{\lambda})]+\lambda[\tempF\,S(\map(\pS)\Vert\map(\qS_{\lambda}))-\tempS S(\pS\Vert\qS_{\lambda})]\\
& \le(1-\lambda)(\tempF-\tempS)S(\qS\Vert\qS_{\lambda})+\lambda(\tempF-\tempS)S(\pS\Vert\qS_{\lambda}) \le0\,.
\end{align*}
The first inequality uses the monotonicity of relative entropy~\cite{muller-hermes_monotonicity_2017}, and the second inequality uses the assumption that $\tempF\le\tempS$. 
Since our derivations holds for all $\pS,\qS,\lambda$, the function $\Wmax$ is concave everywhere.
\end{proof}

\begin{proof}[Proof of \cref{thm:fsupport}]
Suppose that $\qS$ is a maximizer and it does not have full support, and let $\pS$ be any state with full support.
If $\map(\cdot)$ has full support for all inputs,
then $S[\map(\pS)\Vert\map(\qS)]<\infty$. Then, observe that the directional
derivative diverges,  $D_{\pS-\qS}\Wmax({\qS})=\infty$, 
since $\tempS \,S(\pS\Vert\qS)=\infty$ while all other terms in \cref{eq:prop1} are finite. The
strict positivity (actually infinity) of the directional derivative contradicts
the assumption that $\qS$ is a maximizer. Next, consider the case when $\tempS>\tempF$. Then, 
\[
\tempS \,S(\pS\Vert\qS)-\tempF \,S[\map(\pS)\Vert\map(\qS)]\ge(\tempS-\tempF )S(\pS\Vert\qS)=\infty,
\]
where we used the monotonicity of relative entropy~\cite{muller-hermes_monotonicity_2017}. We again have $D_{\pS-\qS}\Wmax({\qS})=\infty$
from \cref{eq:prop1}, contradicting the assumption
that $\qS$ is a maximizer.
\end{proof}

\vfill 
\bibliographystyle{ieeetr}

\bibliography{refs}

\end{document}